%% file: sc.tex
\documentstyle[preprint,epsf]{jpsj} \def\FIGSCALE{16cm}

\def\sf{\rm}

\catcode`\@=11

\newcommand{\bequ}{ \begin{equation} }
\newcommand{\eequ}{ \end{equation} }
\newcommand{\barr}{ \begin{array} }
\newcommand{\earr}{ \end{array} }
\newcommand{\beqarr}{ \begin{eqnarray} }
\newcommand{\eeqarr}{ \end{eqnarray} }

\newcommand{\baralpha}{ \begin{eqnal} \beqarr}
\newcommand{\earalpha}{ \eeqarr \end{eqnal}}

\catcode`\@=12

\begin{document}

\title{%
Critical Exponents of the Metal-Insulator Transition
in the Two-Dimensional Hubbard Model.}

\author
{Nobuo {\sc Furukawa}\footnote{E-mail: furukawa@ginnan.issp.u-tokyo.ac.jp},
 Fakher F. {\sc Assaad}$^1$ \\
 and Masatoshi {\sc Imada}}

\inst{
  Institute for Solid State Physics,
  University of Tokyo, \\
 Roppongi 7-22-1,
  Minato-ku, Tokyo 106\\
$^1$Department of Physics, University of California,\\
 Santa Barbara, CA 93106, U. S. A.
}

\recdate{%
April 15, 1996
}

\kword
{ metal-insulator transition, Hubbard model, hyperscaling,
quantum critical phenomena}

\input abst

\sloppy
\maketitle

\pagebreak

Physics of the metal-insulator (M-I) transition 
has been studied over decades from the point of view of
quantum critical phenomena and strong correlation effects.
Recently, an approach to the M-I transition using a hyperscaling
hypothesis has been developed
by one of the authors (M.I).\cite{Imada95}
Let us consider the filling-controlled
 M-I transition in the ground state, which is controlled by
the shift of the chemical potential
\bequ
  \Delta = \mu - \mu_{\rm c}.
\eequ
Here $\mu$ is the chemical potential while $\mu_{\rm c}$ is
its critical value at the M-I transition.
If we assume the single parameter scaling hypothesis,
the correlation length of the system $\xi$ is described as
\bequ
  \xi \sim |\Delta|^{-\nu},
		\label{HS:xi}
\eequ
and the singular part of the free energy is given by\cite{Hertz76,Golden}
\bequ
 f_{\rm s}(\Delta) \sim |\Delta|^{\nu (d+z)}.
		\label{HS:FreeEne}
\eequ
Here, $z$ is the dynamical exponent, while $d$ is the
dimensionality of the system.
In the metallic phase, the doping concentration $\delta \equiv 1-n$ near
the phase transition is given by
\bequ
  \delta \simeq \frac{\partial f_{\rm s}}{\partial \mu},
\eequ
so that from eq.~(\ref{HS:FreeEne}) we have
\bequ
  \delta \sim
	 |\Delta|^{\nu (d+z) -1} = |\Delta|^{\nu d}.
		\label{HS:delta}
\eequ
Here we used the relation 
$  \nu z = 1 $ which
is obtained from the generalized Josephson relation.

In the case of the Hubbard model on a square lattice near half-filling,
it has been shown by two of the authors 
(N.F. and M.I.)\cite{Furukawa92,Furukawa93} (hereafter NF-MI)
that $\delta$ satisfies
\bequ
  |\Delta| \propto \delta^2.
  \label{QMCmudeltaOld}
\eequ
This result was obtained by computing the chemical potential,
\bequ
\mu = - \frac{ \partial E_{\rm G} } { \partial \delta }
\eequ
at $U/t = 4$ on $4 \times 4$ to $12 \times 12$ lattices,
in the standard notations.
Here $E_{\rm G}$ denotes the ground state energy. The calculations 
were carried out with the zero-temperature 
quantum Monte Carlo algorithm.\cite{Furukawa92,Furukawa93}
This result implies that the 
compressibility diverges as
\bequ
  \kappa \equiv \frac{\partial\delta}{\partial\mu}
     \sim |\Delta|^{-1/2} ,
\eequ
when we approach the M-I transition from the metallic side.
However, in NF-MI, an estimate of the statistical error
of the exponent was not done, mainly  due to large statistical uncertainty
in location of the critical point, $\mu_{\rm c}$.  

Recently, two of the authors (F.F.A. and M.I.)\cite{Assaad96a,Assaad95xb}
(hereafter FFA-MI) have computed with QMC methods, zero temperature
imaginary time displaced Green function in the insulating phase. From this
data they obtain an accurate estimate of the critical chemical potential
at $U/t = 4$: $\mu_{\rm c} = 0.67 \pm 0.05$ in units of the 
hopping matrix element. This result is obtained by extrapolating to
the thermodynamic limit from data on $ 4\times 4 $ to 
$ 16 \times 16 $ lattice.

Combining this value with the chemical potential
data of NF-MI in the metallic phase leads to:
\bequ
    | \Delta | \propto \delta^{2.38\pm 0.43} 
\eequ 
(see Fig. 1). 
From this result and the assumption of 
hyperscaling we obtain a value of the correlation length exponent
\bequ
\nu_{\kappa} = 0.21 \pm 0.04.
\eequ

\begin{figure}[hb]
\epsfxsize=\FIGSCALE\epsfbox{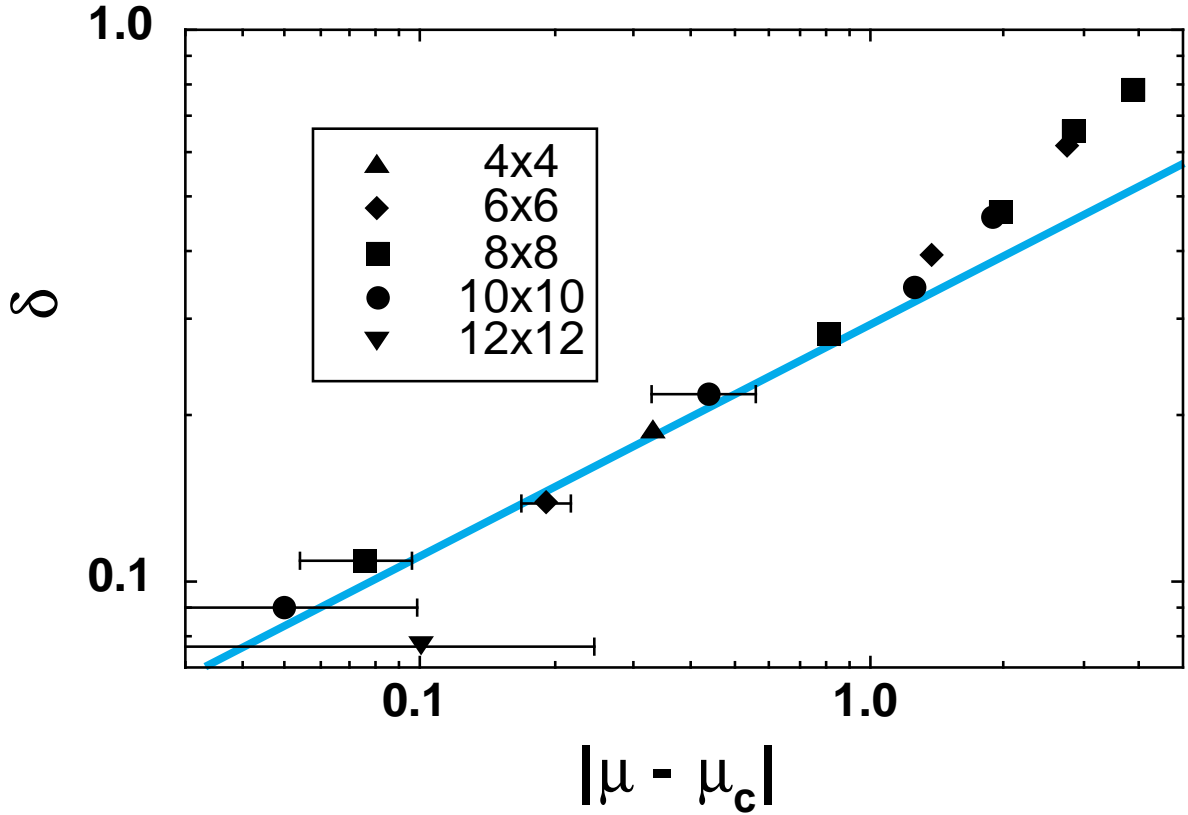}
\caption{ Log-log plot of $|\mu - \mu_{\rm c} |/t$ verses $ \delta$ in the
metallic phase. Here, $\mu_{\rm c}/t = 0.67 \pm 0.05$. The line
is the least-squares fit to the data.}
\label{FigMetal}
\end{figure}

In order to directly determine the  correlation length exponent, 
FFA-MI\cite{Assaad95xb}
have computed the Green function $G(\mib{r},\omega= \mu)$ for values of
the chemical potential within the charge gap.  For those values of the
chemical potential,  
$G(\mib{r},\omega= \mu) \propto e^{-|\mib{r}|/\xi_l}$.
 The localization length, $\xi_l$, 
diverges as the M-I transition is
approached from the insulating side. We obtain numerically:
\bequ
    \xi_l \propto | \Delta | ^{-0.26 \pm 0.05 }
\eequ
(see Fig. 2). From this result, we obtain an estimate of the
correlation length exponent:
\bequ
  \nu_l = 0.26 \pm 0.05.
\eequ

\begin{figure}[htb]
\epsfxsize=\FIGSCALE\epsfbox{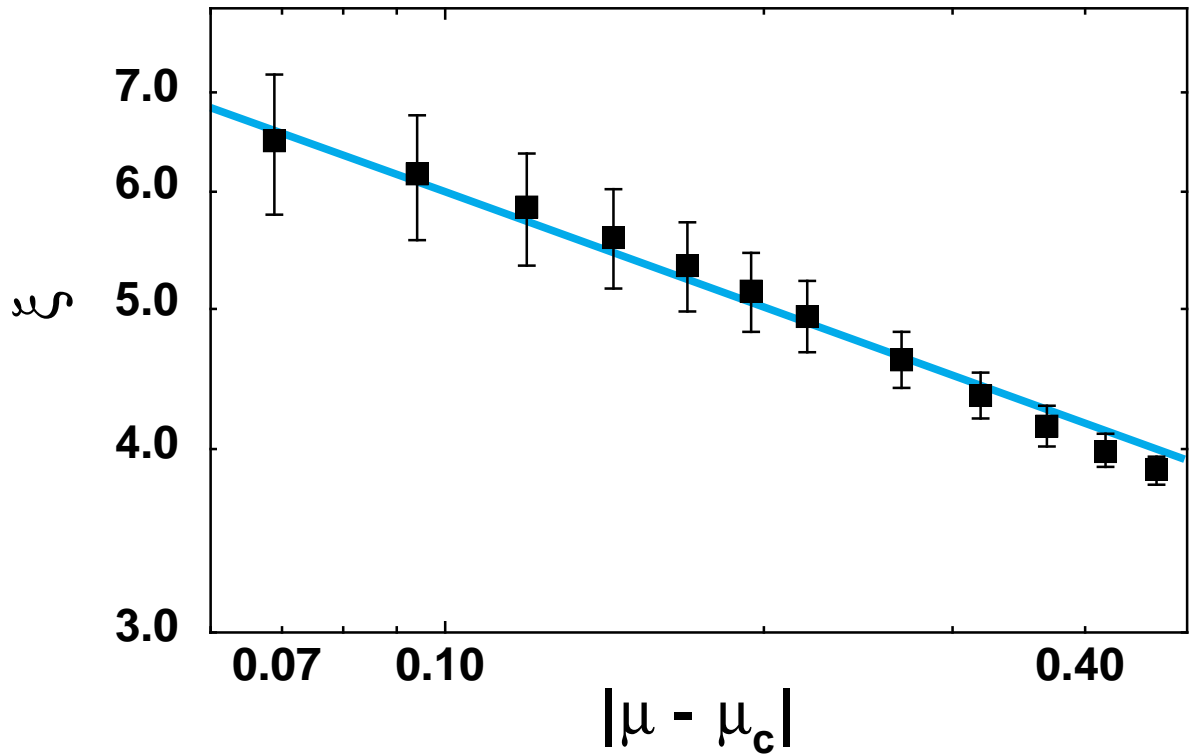}
\caption{Log-log plot of $ |\mu - \mu_{\rm c} |/t$ verses $\xi_l$ in the
insulating phase.  Here, $\mu_{\rm c}/t = 0.67 \pm 0.05$. The line
is the least-squares fit to the data.}
\label{FigInsl}
\end{figure}

If hyperscaling is valid, one expects the relation:
\bequ
\nu_{\kappa} \equiv \nu_l.
	\label{HS:nu-equiv}
\eequ
That the exponents should be equal on  either side 
of the critical point 
is obtained  through the scaling properties,  and from the
fact that the insulating and metallic phases may be connected
smoothly around the critical 
point $\Delta=0$.\cite{Golden,Note1}
The main result of this paper is that eq.~(\ref{HS:nu-equiv}) is
satisfied within statistical uncertainty, 
thus putting on a firm numerical basis the assumption of 
hyperscaling. The M-I transition in 2D Hubbard model belongs to the
universality class characterized by $\nu = 1/4$ and $z=4$. This
stands in contrast to the generic band M-I transition in all
dimensions as well as to the Mott transition in the 1D Hubbard model,
which both belong to the universality class $\nu=1/2$ and $z=2$.
 The M-I transition in 2D Hubbard model belongs to the
universality class characterized by $\nu = 1/4$ and $z=4$.\cite{Abraham}
 This stands in contrast to the generic band M-I transition in all
dimensions as well as to the Mott transition in the 1D Hubbard model,
which both belong to the universality class $\nu=1/2$ 
and $z=2$.\cite{Continentino}

The authors are grateful to T. Ogitsu for help in the use of
supercomputers.
The numerical calculation was partially performed on the
 FACOM VPP500 at the Supercomputer Center, 
Inst.\ for Solid State Phys., Univ.\ of Tokyo.
This work is supported by a Grant-in-Aid for
Scientific Research on the Priority Area `Anomalous Metallic
State near the Mott Transition' from the Ministry of Education,
Science and Culture, Japan.

\pagebreak

\end{document}

%% file: abst.tex
\abst{
We study the filling-controlled 
metal-insulator transition in the two-dimensional Hubbard model 
near half-filling
with the use of zero temperature quantum Monte Carlo methods. 
In the metallic phase, the compressibility behaves as
$\kappa \propto  |\mu - \mu_c| ^{-0.58 \pm 0.08}$  where $\mu_c$ is 
the critical chemical potential.
In the insulating phase, the localization length follows
$\xi_l \propto |\mu - \mu_c|^{-\nu_l}$ with $\nu_l = 0.26 \pm 0.05$.  
Under the assumption of hyperscaling, the
compressibility data leads to a correlation length exponent
$\nu_\kappa = 0.21 \pm 0.04$. 
Our results show that the exponents $\nu_\kappa$  and $\nu_l$
agree within statistical uncertainty.
This confirms the assumption of hyperscaling with correlation length 
exponent $\nu = 1/4$ and dynamical exponent $z = 4$.
In contrast the metal-insulator transition in the generic band 
insulators in all dimensions as well as in the one-dimensional 
Hubbard model satisfy the hyperscaling assumption with exponents 
$\nu = 1/2$ and  $z = 2$. }